\documentclass[twocolumn,showpacs,prl]{revtex4}
\usepackage{bm}
\usepackage{graphicx}
\usepackage{amsmath}
\newcommand{\nix}[1]{}

\begin{document}

\title{
Valley-Orbit Photocurrents in (111)-oriented Si-MOSFETs}
\author{J.~Karch,$^1$ S.~A.~Tarasenko,$^2$ E.~L.~Ivchenko,$^2$
J.~Kamann,$^1$ P.~Olbrich,$^1$ M.~Utz,$^1$ Z.~D.~Kvon,$^3$ and S.~D.~Ganichev$^{1}$}

\affiliation{$^1$Terahertz Center, University of Regensburg,
93040 Regensburg, Germany}
\affiliation{$^2$Ioffe Physical-Technical Institute, Russian Academy of
Sciences, 194021 St.\,Petersburg, Russia}
\affiliation{$^3$ Institute of Semiconductor Physics, Russian Academy
of Sciences, 630090 Novosibirsk, Russia}

\begin{abstract}
We demonstrate the injection of pure valley-orbit currents in multi-valley semiconductors
and present the theory of this effect. 
We studied photo-induced transport in $n$-doped (111)-oriented silicon metal-oxide-semiconductor field effect transistors
at room temperature. 
By shining circularly polarized light on exact oriented structures with six equivalent valleys,
non-zero electron fluxes within each valley are generated, 
which compensate each other and do not yield a net electric current. 
By disturbing the balance between the valley fluxes, 
in this work by applying linearly polarized radiation as well as by introducing a nonequivalence of the valleys by disorientation,
we approve that the pure valley currents can be converted into a measurable electric current.
%
%
\end{abstract}

\date{\today}

\maketitle


Free carriers in solids can carry both positive and negative electric charge, which builds the basis for bipolar electronics.
In addition, they possess degrees of freedom related to spin or valley
degeneracy, the latter being relevant in multi-valley semiconductors. 
The degeneracies enable one to engineer various distributions
of carriers in the momentum, spin, and valley spaces.
A controllable way of occupying a particular spin state or filling a particular valley is a key
ingredient, respectively, for spintronics~\cite{spintronics} 
or valleytronics~\cite{TarIvch05,beenakker1} aimed at the development of novel  solid-state devices. 
The candidates appropriate for the realization of valleytronics concepts include multi-valley semiconductors such as
silicon~\cite{TarIvch05,mcfarland}, graphene~\cite{beenakker1,garsia,Moskalenko}, and carbon nanotubes~\cite{TarIvch05}.
A remarkable selective population of the electron states is achieved by optical
pumping with polarized light: circularly 
for spin polarization~\cite{spin_physics} and linearly polarized for
``valley polarization''~\cite{sokolov}. 
Besides the selective
valley population, the radiation can also cause a particle flux
${\bm i}_{\nu}$ within each valley, whose direction and magnitude
depend on the valley number $\nu$. 
In general, the total electric current ${\bm j} = e \sum_{\nu}
{\bm i}_{\nu}$ is non-zero, measurable by conventional electric methods. 
However, for special geometries and light polarizations,
the partial fluxes
${\bm i}_{\nu}$ do exist but the total electric current vanishes.
This is the pure valley current proposed in
Ref.~\cite{TarIvch05} and implying the device's potential use of the valley index
in analogy to the pure spin currents in spintronics and topological electronics~\cite{sipe1,naturephys,sipe2,TarIvch08,Kane,Koenig}. 

\begin{figure}[t]
\includegraphics[width=0.9\linewidth]{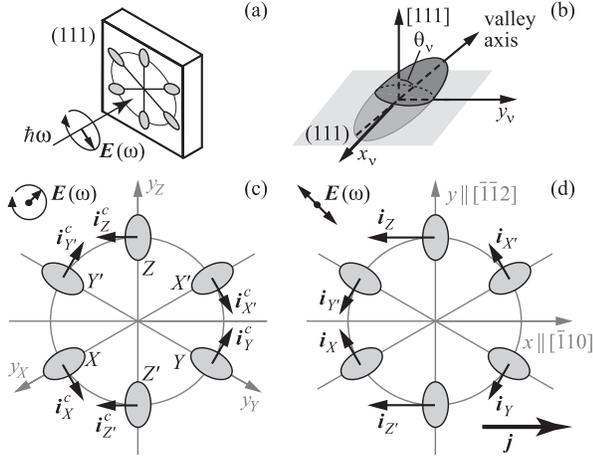} \caption{
(a) Excitation of (111)-oriented Si-MOS structures
with, e.g., circularly polarized light.
(b) Orientation of an individual valley.
(c)
Distribution of helicity-dependent fluxes 
over valleys induced in the exact (111) Si-MOSFET under normal incidence of circularly polarized light. 
The fluxes ${\bm i}^c_{\nu}$ are directed perpendicularly to the valley principal axes and compensate each other to nullify the net electric current.
The in-plane coordinates $ y_{\nu}$ attached to the valleys $X$, $Y$ and $Z$ are shown.
(d)
Intravalley fluxes under excitation with linearly polarized light,
the total current ${\bm j} = e \sum_{\nu} {\bm i}_{\nu}\neq 0$. 
} \label{figure1}
\end{figure}

Here, we report on the vivid evidence of the existence of pure valley-orbit currents
and demonstrate that the individual control of electron fluxes in valleys can be achieved 
by the excitation of (111)-oriented Si-metal-oxide-semiconductor (Si-MOS) structures with polarized light.
Such two-dimensional systems contain six equivalent 
electron valleys~\cite{mcfarland,Si111}, Fig.~\ref{figure1}, 
and possess the overall point-group symmetry C$_{3v}$, i.e., 
they are invariant with respect to the rotation by an angle of $120^\circ$.
The irradiation of (111)-oriented Si-MOS structures leads to an emergence of
the fluxes ${\bm i}_{\nu}$ in all six valleys, Fig.~\ref{figure1}(a). 
The generation of the fluxes ${\bm i}_{\nu}$ is microscopically caused by the low symmetry of individual valleys,
Fig.~\ref{figure1}(b),
and stems from asymmetric electron photoexcitation. 
This mechanism is similar to those considered in Refs.~\cite{TarIvch05,Magarill89,JPCM},
but out of scope of this Letter and will be described elsewhere.
Instead, we focus on the distribution of the fluxes $\bm{i}_{\nu}$ over the valleys
forming the pure orbital currents or, under certain conditions, an electric current. 
The symmetry analysis shows that circularly polarized light
normally incident upon the sample induces
helicity-sensitive intravalley fluxes ${\bm i}_{\nu}^c$ in the
directions perpendicular to the principal valley axes resulting in a zero net photocurrent, see Fig.~\ref{figure1}(c). 
On the other hand, the
theory allows the generation of a total electric current by
normally incident and linearly polarized radiation, Fig.~\ref{figure1}(d).
Thus, a first experimental evidence 
of valley currents in (111) Si-MOS structures is
the lack of a net photocurrent for circularly polarized
light and its observation for linear polarization. 
A further confirmation is the appearance of a 
photocurrent in samples with broken equivalence of valleys, e.g., in structures with
the surface normal $\bm n$ slightly tilted from the [111] axis.

\begin{figure}[t]
\includegraphics[width=0.81\linewidth]{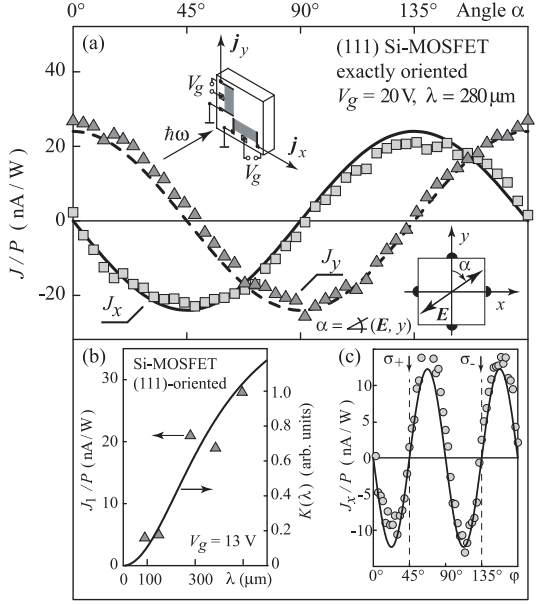}
\caption{(a) Photocurrent as function of the azimuth angle $\alpha$ measured
in the precisely (111)-oriented MOSFET in the
crystallographic directions $x$ and $y$
for a gate voltage of $V_g = 20$\,V and the
wavelength $\lambda$~=~280~$\mu$m.
Lines are fits to Eqs.~(\ref{sincos})
with one fitting parameter $\chi$.
The upper inset shows the experimental geometry,
the lower the definition of the angle $\alpha$ with respect
to the sample orientation.
(b) Comparison of the wavelength dependencies of the linear photocurrent
$J_1 =  [J_y(\alpha=0^\circ)-J_y(\alpha=90^\circ)]/2 \propto \chi$
measured for $V_g=13$\,V (triangles) and the calculated coefficient of free carrier absorption (solid curve).
(c) Photocurrent in the $x$ direction as function of the angle $\varphi$,
full line is a fit to Eq.~(\ref{jxpprime}).}
 \label{figure2}
\end{figure}

To verify the theoretical predictions,
we have experimentally studied photocurrents in MOS field-effect
transistors (MOSFETs) fabricated on exact and miscut (111) silicon surfaces.
The studied Si samples with the surface precisely oriented along
the (111) plane contain two transistors, see inset of
Fig.~\ref{figure2}(a), each prepared with a channel length of
$1.5$\,mm, a width of $0.5$\,mm and the channel directions
parallel either to $x \parallel [\bar{1} 1 0]$ or $y \parallel [\bar{1} \bar{1} 2]$. 
The channels are covered by $100$\,nm thick SiO$_2$ layers and
semitransparent polycrystalline Si gates.
The variation of the gate voltage $V_g$ from $1$ to
$20$\,V changes the
carrier density $N_s$ from
$2.0 \times 10^{11}$ to $4.1 \times 10^{12}$\,cm$^{-2}$. As a
reference sample we also fabricated MOSFETs on exact (001)
Si surfaces with similar characteristics.

\begin{figure}[t]
\includegraphics[width=0.81\linewidth]{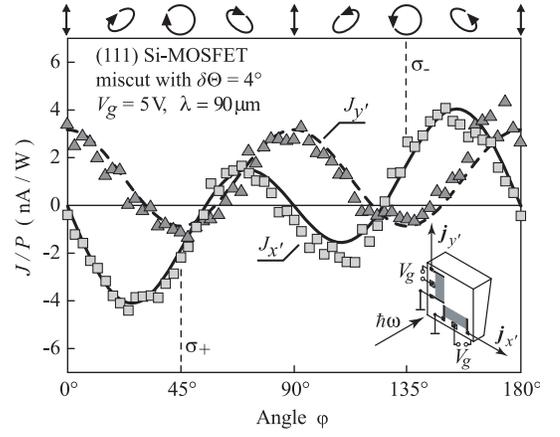}
\caption{Photocurrent as a function of the angle $\varphi$ measured
in the transistors prepared on the miscut surface for $V_g=5$\,V and
radiation with the wavelength $\lambda$~=~90~$\mu$m.
Lines are fits to the phenomenological
Eqs.~(\ref{fitting}).
The inset shows the experimental geometry.
On top the polarization ellipses corresponding to various angles
$\varphi$ are illustrated.
} \label{figure3}
\end{figure}

Another set of MOSFETs is fabricated on Si surfaces with the
normal ${\bm n}$ rotated from the $(111)$ orientation around the
axis $x'= x \parallel [\bar{1} 1 0]$ by an angle of $ \delta \Theta =
4^\circ$. The point-group symmetry of this system is C$_s$ consisting of the identity 
element and the mirror rotation plane ($1 \bar{1} 0$). 
Two transistors with the same size are prepared on the miscut substrate with the channels
oriented either along $x^\prime$ or along the inclination
axis $y' \perp {\bm n}, x'$ (see inset in Fig.~\ref{figure3}).
They have $70$\,nm thick SiO$_2$ layers and semitransparent polycrystalline Si gates.
Variation in 
$V_g$ from $1$ to $20$\,V changes 
$N_s$ from $3.0 \times 10^{11}$ to $5.7 \times 10^{12}$\,cm$^{-2}$.

All kinds of MOSFETs used in the experiment exhibit
room-temperature electron mobilities $\mu$ in the channel
between $400$ and $800$\,cm$^2$/Vs. The orientation of the Si
surfaces has been proved applying an X-ray diffractometer. The
accuracy of the orientation is better than $0.5^\circ$.

For optical excitation we use the emission of an optically pumped terahertz (THz)
molecular gas laser~\cite{GanichevPrettl}.
The radiation with a pulse length of 100~ns  and a peak power
$P \sim $10~kW is obtained at the wavelengths $\lambda=$ 90,
148, 280, 385 and 496~$\mu$m corresponding to photon energies $\hbar \omega$ between $13.7$ and $2.5$\,meV. 

The experimental geometries are
illustrated in the insets of Figs.~\ref{figure2}
and~\ref{figure3}. All experiments are performed at normal incidence of radiation and room temperature. 
In this setup, the THz radiation causes indirect Drude-like optical
transitions in the Si-MOSFETs.
Various polarization states of the radiation are achieved by
transmitting the laser beam, which is initially linearly polarized along the $y$ (or $y'$) axis,
through $\lambda$/2 or $\lambda$/4 crystal quartz plates. 
By applying the $\lambda/2$ plates, 
we vary the azimuth angle $\alpha$ between the polarization plane of the radiation incident upon the sample
and the $y$ (or $y^\prime$) axis,
see the inset of Fig.~\ref{figure2}(a). 
By applying $\lambda$/4 plates we transfer linearly  
to elliptically polarized radiation. 
In this case, the polarization state is determined by the angle $\varphi$ between the $y$ (or $y^\prime$) axis and the optical axis of the plate.
The Stokes parameters ${\cal S} = e_x e_y^* + e_y e_x^*$, ${\cal C} = |e_x|^2 - |e_y|^2$ and
the degree of circular polarization $P_{\rm circ}$,
defined by $i [\bm{e} \times \bm{e}^*] = (\bm{q}/q)P_{\rm circ}$,
depend on the angles $\alpha$ and $\varphi$ as
\begin{equation} \label{Stokes}
{\cal S}(\alpha) = \sin{2 \alpha} \:, \;\;
{\cal C}(\alpha) = - \cos{2 \alpha} \:, 
\end{equation}
\[
{\cal S}(\varphi) = \frac12 \sin{4\varphi}\:,\;\;
{\cal C}(\varphi) = - \cos^2{2\varphi}\:, \;\; P_{\rm circ} = \sin {2 \varphi} \:.
\]
Here, ${\bm e}$ is the polarization unit vector ${\bm E}/|{\bm E}|$ and $\bm{q}$ is the photon wave vector directed
along $-z$.
The incident polarization states are sketched for characteristic angles $\varphi$ on top of
Fig.~\ref{figure3}. 
The photocurrents are measured between 
source and drain contacts of the unbiased transistors via the
voltage drop across a 50~$\Omega$ load resistor.


Irradiating the exact (111) Si-MOSFETs with linearly
polarized THz radiation at normal incidence generates
photocurrent signals with a temporal shape reproducing the laser
pulse.
The 
dependence on
the azimuth angle $\alpha$ is shown in Fig.~\ref{figure2}(a), the
data are 
well fitted by 
\begin{equation} \label{sincos}
J_x (\alpha) =  \chi \sin{2 \alpha}\:,\:J_y (\alpha) = - \chi \cos{2 \alpha}
\end{equation}
with the same prefactor $\chi = - 24$ nA/W, in agreement with the theoretical expectation for the point group C$_{3v}$.
It has been checked that, in the
reference MOSFETs prepared on the exact (001) Si surface,
normal-incidence photocurrents are absent for any polarization of
radiation~\cite{JPCM}.
Triangles in Fig.~\ref{figure2}(b) show that, as expected for the free-carrier absorption, 
the signal rises with increasing the wavelength following the Drude formula of high-frequency conductivity~\cite{GaN2008,Tarasenko10}. 
%

The variation of the photocurrent with the angle $\varphi$ presented
in Fig.~\ref{figure2}(c) clearly demonstrates that the current
vanishes for circularly polarized photoexcitation, $\sigma_+$ and
$\sigma_-$, realized at $\varphi = 45^{\circ}$ and $135^{\circ}$.
The best fit of the dependencies $J_x(\varphi)$ and $J_y(\varphi)$ (not shown) 
contains no contributions from $P_{\rm circ}$ and reduces to 
\begin{equation} \label{jxpprime}
J_x(\varphi) = \frac{\chi}{2} \sin 4 \varphi \:,\;\;\; J_y(\varphi) = - \chi \cos^2 2 \varphi \:,
\end{equation}
with the same coefficient $\chi$ as in Eq.~(\ref{sincos}).


The symmetry considerations for the miscut (111) Si transistors with the C$_s$ point group
lead to the following phenomenological equations for the 
in-plane photocurrent components as a function of the angle $\varphi$
%
%
%
\begin{equation} \label{fitting}
J_{x'}  = \gamma P_{\rm circ} + \frac{\chi_1}{2} \sin 4 \varphi \:,\;\;
J_{y'}  = \chi_2 - \chi_3 \cos^2 2 \varphi \:. 
\end{equation}
%
Figure \ref{figure3} shows the effect of a miscut angle $\delta \Theta$. 
The curves in the figure are well fitted by Eqs.~(\ref{fitting}) with the values of coefficients (in nA/W):
$\gamma = - 1.8$, $\chi_1 = - 5.5$, $\chi_2 = - 0.8$, and $\chi_3 = - 4.0$.
As compared with the exact (111) Si sample, the inclination leads to three distinctive features:
(i) a current along the $x'$ axis proportional to the radiation helicity $P_{\rm circ}$ is generated under normal incidence,
(ii) the current component $J_{y'}$ acquires a polarization-independent contribution, and 
(iii) the absolute values of the coefficients $\chi_1, \chi_3$
describing the currents induced by linearly polarized light become different.

Now we show that the observed dependencies~(\ref{sincos})-(\ref{fitting}) 
are readily obtained by summing up the photocurrents $\bm{j}_{\nu} = e \bm{i}_{\nu}$ induced in individual valleys.
Each valley $\nu$ ($\nu=X,Y,Z,X',Y'$ and $Z'$, see Fig.~\ref{figure1}) has only one non-trivial symmetry element, 
the reflection $x_{\nu} \rightarrow - x_{\nu}$, 
described by the point group $C_s$. 
Here, we use the following in-plane frame: 
the axis $x_{\nu}$ is perpendicular to the plane containing the principal valley axis and the channel normal $\bm{n}$,
and the axis $y_{\nu}$ lies in this plane. 
Particularly, for ${\bm n}
\parallel [111]$, the axes $x_{Z}$ and $y_{Z}$ coincide
with $[\bar{1} 1 0]$ and $[\bar{1} \bar{1} 2]$, respectively.
The current density emerging in the valley $\nu$ is phenomenologically given by
\begin{eqnarray} \label{singleW}
j^{(\nu)}_{x_{\nu}}/I &=& A (\theta_{\nu}) P_{\rm circ} + B (\theta_{\nu}) ( e_{x_{\nu}} e^*_{y_{\nu}} +
e_{y_{\nu}} e^*_{x_{\nu}}) \:,\\
j^{(\nu)}_{y_{\nu}}/I &=& C (\theta_{\nu}) + D (\theta_{\nu}) (|e_{x_{\nu}}|^2 - |e_{y_{\nu}}|^2)]\:, \nonumber
\end{eqnarray}
where $I$ is the light intensity,
$A(\theta_{\nu})$, $B(\theta_{\nu})$, $C(\theta_{\nu})$, and $D(\theta_{\nu})$
are polarization-independent
coefficients, $\theta_{\nu}$ is the angle between the normal $\bm{n}$ and the principal axis of the valley $\nu$,
$e_{x_{\nu}}$ and $e_{y_{\nu}}$ are the projections of the polarization vector ${\bm e}$ on the $x_{\nu}$ and $y_{\nu}$ axes, respectively.

The total electric current is a sum of the single-valley contributions
${\bm j}^{(\nu)}$. In fact, the contributions due to the valleys
$X, X'$ (or $Y,Y'$ and $Z,Z'$) coincide so that the total photocurrent simplifies to
\begin{equation} \label{sum}
{\bm j} = 2 \sum\limits_{\nu = X,Y,Z} {\bm j}^{(\nu)}\:.
\end{equation}
By projecting all the partial vectors ${\bm j}^{(\nu)}$ on the transistor axes $x$ and $y$ (or $x'$ and $y'$)
we obtain
\begin{eqnarray} \label{xzyz}
j_{x} = 2 \sum\limits_{\nu = X,Y,Z} \left[ j^{(\nu)}_{x_{\nu}}  \cos{\phi_{\nu}} 
- j^{(\nu)}_{y_{\nu}} \sin{\phi_{\nu}} \right] \:,\\
j_{y} = 2 \sum\limits_{\nu = X,Y,Z} \left[ j^{(\nu)}_{x_{\nu}} \sin{\phi_{\nu}}
+ j^{(\nu)}_{y_{\nu}} \cos{\phi_{\nu}} \right] \:, \nonumber
\end{eqnarray}
where $\phi_{\nu}$ is the angle between the axes $x_{\nu}$ and $x$. 

In the exact (111) Si-MOS structure, all valleys are equivalent, 
$\theta_{\nu}=\Theta_0$ with $\cos \Theta_0 =1/\sqrt{3}$, and the in-plane angles $\phi_{\nu}$ are given by $\phi_{X} = - \phi_{Y} = 120^{\circ}$, $\phi_{Z}=0$. Therefore, we derive 
\begin{eqnarray}\label{totalcurrent}
j_{x}/I = 3 (B_0+D_0) (e_x e_y^* + e_y e_x^*) \:,\\
j_{y}/I = 3 (B_0+D_0) (|e_x|^2 - |e_y|^2) \:, \nonumber
\end{eqnarray}
where $B_0 = B(\Theta_0)$ and $D_0 = D(\Theta_0)$. Equation~(\ref{totalcurrent})
exactly reproduces the experimentally observed polarization dependencies~(\ref{sincos})-(\ref{jxpprime})
with the fitting parameter $\chi$ given by $3(B_0+D_0)$. 
Thus, we conclude that, while both circularly polarized and unpolarized radiation give rise to an electric currents $\bm{j}_{\nu}$ in all valleys, the total photocurrent can only be excited by linearly polarized light.

To describe the photocurrent in miscut structures, 
we introduce the polar angles $\Theta$ and $\Phi$ of the normal ${\bm n}$ in the Cartesian system [100], [010], [001].
If the channel normal $\bm{n}$ is close to $[111]$, then the angles $\Theta$ and $\Phi$ are close to $\Theta_0$ and $\Phi_0=45^\circ$, respectively. 
By expanding $\phi_{\nu}$, $\theta_{\nu}$ in powers of $\delta \Theta = \Theta - \Theta_0$ and $\delta \Phi = \Phi - \Phi_0$
one can derive the expression for the photocurrent in miscut structures.
In the approximation linear in $\delta \Theta$ and $\delta \Phi$,
the total helicity-dependent current reduces to
\begin{equation} \label{eq1}
{\bm j}^c /I = Q P_{\rm circ} \, {\bm l} \times {\bm n} \:,
\end{equation}
where $Q = 3 \left( A_0/\sqrt{2} + A^{\prime}_0 \right)$, $A^{\prime}_0= dA/d\theta |_{\theta=\Theta_0}$, and ${\bm l}$ is the unit vector along the crystallographic axis $[111]$. 
While deriving Eq.~(\ref{eq1}) we took into account that 
(i) for a fixed value of the total electron density, the electron chemical potential is independent of $\delta \Theta$
and $\delta \Phi$ 
(ii) the angles $\theta_{\nu}$ are expressed in terms of $\delta \Theta$, $\delta \Phi$ by $\theta_Z = \Theta_0 + \delta \Theta$
and $\theta_{X,Y} = \Theta_0 - \delta \Theta/2 \pm \delta \Phi/\sqrt{2}$, and
(iii) the components $({\bm l} \times {\bm n})_{x'}$ and $({\bm l} \times {\bm n})_{y'}$
are given by $\delta \Theta$ and $\sqrt{2/3} \, \delta \Phi$, respectively. 
Equation~(\ref{eq1}) shows that the helicity-dependent photocurrent excited by normally incident radiation appears only in miscut structures and in the direction perpendicular to the inclination axis. This is the behavior that is observed in all Si-MOSFETs studied in experiment, see Figs.~\ref{figure2}(c) and~\ref{figure3}.
It follows from the experimentally measured value of the ratio $\gamma/\chi_3$ and from Eq.~(\ref{eq1}) that the ratio 
$A_0/\sqrt{2}+A'_0$ and $B_0 + D_0$ amounts 5. An estimation based on the microscopic theory~\cite{PRBSi,Tarasenko10} of orbital mechanisms of dc currents driven by ac electric fields gives for the latter ratio a value of unity. This mismatch needs a further analysis.

The valley currents with zero net charge transfer can be classified according to their behavior under the symmetry operations. 
In (111)-oriented Si-MOS structures of the $C_{3v}$ point group, the pure valley currents can belong to the irreducible representations $A_1$, $A_2$ and $E$.
The scalar representation $A_1$ describes the pure valley current $j_{\rm valley}^{(A_1)} = \sum_{\nu} j_{y_{\nu}}^{(\nu)}$;
this current is excited by normally incident unpolarized radiation. 
The pseudoscalar representation $A_2$ describes the valley current $j_{\rm valley}^{(A_2)} = \sum_{\nu} j_{x_{\nu}}^{(\nu)}$, 
which changes its sign upon reflection in any of the three mirror planes of the point group $C_{3v}$. 
The contribution $j_{\rm valley}^{(A_2)}$ is induced by circularly polarized radiation, see Fig.~\ref{figure1}(c),
and reverses its sign upon inversion of the photon helicity. 
Finally, the two-dimensional vector representation $E$ describes the valley current with the components
$j_{{\rm valley},x}^{(E)} = \bigl(2 j_{x_Z}^{(Z)} - j_{x_X}^{(X)} - j_{x_Y}^{(Y)} \bigr) + \sqrt{3} \bigl( j_{y_X}^{(X)} - j_{y_Y}^{(Y)}\bigr) $ and $j_{{\rm valley},y}^{(E)} = \bigl( - 2 j_{y_Z}^{(Z)} + j_{y_X}^{(X)}+j_{y_Y}^{(Y)}\bigr) + \sqrt{3} \bigl(j_{x_X}^{(X)} - j_{x_Y}^{(Y)}\bigr) $. 
The polarization dependence of the pure valley current $\bm{j}_{{\rm valley}}^{(E)}$ excited by normally incident radiation is given by Eq.~(\ref{totalcurrent}) where $D_0$ is replaced by $-D_0$. 

In conventional electric measurements, the pure valley currents are hidden and special methods are needed to reveal them. 
A possible experiment to observe 
$j_{\rm valley}^{(A_2)}$ 
is the generation of the second harmonic.
Indeed, under normal incidence of the probe light, the dielectric polarization at the second harmonic frequency is given by
\begin{eqnarray}
P^{(2)}_x &=& \chi^{(2)}_{xxy} 2 {\cal E}_x {\cal E}_y + \chi^{(3)}_{xxx} \, j_{\rm valley}^{(A_2)}
({\cal E}_x^2 - {\cal E}^2_y)\:, \nonumber\\ 
P^{(2)}_y &=&  \chi^{(2)}_{xxy} ({\cal E}_x^2 - {\cal E}^2_y) - \chi^{(3)}_{xxx} \, j_{\rm valley}^{(A_2)} \, 2 {\cal E}_x {\cal E}_y\:, \nonumber
\end{eqnarray} 
where ${\cal E}$ is the probe electric-field amplitude, $\chi^{(2)}_{xxy} = \chi^{(2)}_{yxx} = - \chi^{(2)}_{yyy}$ the second-order susceptibility of the equilibrium system and $\chi^{(3)}_{xxx}= - \chi^{(3)}_{xyy} = - \chi^{(3)}_{yxy}$ the third-order susceptibility, the second order in ${\cal E}$ and the first in $j_{\rm valley}^{(A_2)}$.
Note that the effect of the gate voltage on the generation of the second harmonic has been previously observed in (001) Si-MOSFETs~\cite{Second}. Another possibility is the {\it orbital} Kerr or Faraday rotation of the probe beam caused by the pure orbit-valley current with the rotation angle $\Phi \propto j_{\rm valley}^{(A_2)}$, similar to 
spin Kerr or Faraday effect.


Support from DFG (SPP1285), Linkage Grant of IB of BMBF at DLR, 
RFBR, Russian Ministry of Education and Sciences, 
and ``Dynasty'' Foundation ICFPM is gratefully acknowledged.

%
%



\end{document}